\documentclass[a4paper,10pt]{article}

\usepackage{amsmath}
\usepackage{amsfonts}
\usepackage{amssymb}
\usepackage{graphicx}
\input{epsfx}

\title{The Inert Doublet Model and evolution of the Universe}
\author{Maria Krawczyk\footnote{Talk presented at \textit{10th Hellenic School on Elementary Particle Physics
and Gravity, Corfu 2010}. Proceedings are to be published in Fortschritte der Physik.} \, and Dorota Soko\l owska \\
\textit{University of Warsaw, Faculty of Physics, Hoza 69, 00-681 Warsaw, Poland}}

\begin{document}

\maketitle

\begin{abstract}

Inert Doublet Model (IDM) is a minimal extension of the Standard Model with the second scalar doublet that may provide a Dark Matter candidate. In this paper we consider the different variants of the evolution of the Universe after inflation, that lead towards the Inert phase today.

\end{abstract}

\section{Introduction}

\paragraph{2HDM}
The Two Higgs Doublet Model (2HDM) is a minimal extension of the Standard Model (SM) by the second scalar doublet \cite{Lee:1973iz}. The electroweak symmetry breaking (EWSB) via the Brout-Englert-Higgs-Kibble (BEHK) mechanism is described by the Lagrangian
\begin{equation}
{ \cal L}={ \cal L}^{SM}_{ gf } +{ \cal L}_H + {\cal L}_Y \,, \quad { \cal L}_H=T-V\, .
\label{lagrbas}
\end{equation}
${\cal L}^{SM}_{gf}$ is the $SU(2)\times U(1)$ Standard Model
interaction of gauge bosons and fermions.
In 2HDM the scalar Lagrangian of two SU(2) doublets $\Phi_{S,D}$, with weak hypercharge Y=+1, consists of the standard kinetic term $T$ and  
the potential $V$ of general form:
\begin{eqnarray}
V & = &
\frac{\lambda_{1}}{2} \left(\Phi_{S}^{\dagger} \Phi_{S}\right)^{2} + \frac{\lambda_{2}}{2} \left(\Phi_{D}^{\dagger} \Phi_{D}\right)^{2}  +\lambda_{3} \left(\Phi_{S}^{\dagger} \Phi_{S}\right) \left(\Phi_{D}^{\dagger} \Phi_{D}\right) + \lambda_{4} \left(\Phi_{S}^{\dagger} \Phi_{D}\right) \left(\Phi_{D}^{\dagger} \Phi_{S}\right) \nonumber \\
& & + \left[ \frac{1}{2} \lambda_{5} \left(\Phi_{S}^{\dagger} \Phi_{D}\right) \left(\Phi_{S}^{\dagger} \Phi_{D}\right) + \lambda_{6} \left(\Phi_{S}^{\dagger} \Phi_{S}\right) \left(\Phi_{S}^{\dagger} \Phi_{D}\right) +  \lambda_{7} \left(\Phi_{D}^{\dagger} \Phi_{D}\right) \left(\Phi_{S}^{\dagger} \Phi_{D}\right) + H.c. \right] \nonumber \\
& &  - \frac{1}{2}m_{11}^{2} \left(\Phi_{S}^{\dagger} \Phi_{S}\right) -\frac{1}{2}m_{22}^{2} \left(\Phi_{D}^{\dagger} \Phi_{D}\right) -\frac{1}{2}m_{12}^{2} \left(\Phi_{S}^{\dagger} \Phi_{D}\right) -\frac{1}{2}(m_{12}^{2})^{*} \left(\Phi_{D}^{\dagger} \Phi_{S}\right), \label{V}
\end{eqnarray}
where $\lambda_{1-4}, m_{11}^2, m_{22}^2 \in \mathbb{R}$ and $\lambda_{5-7}, m_{12}^2 \in \mathbb{C}$. If the explicit $Z_2$-symmetry is present in the potential then the soft violating term $\propto m_{12}^2$ and hard violating terms $\propto \lambda_{6,7}$ are absent.

 ${\cal L}_Y$ describes the Yukawa interaction of SM fermions $\psi_f$. We will assume Model I, meaning that only one scalar doublet $\Phi_S$ interacts with fermions, and ${\cal L}_Y (\psi_f,\Phi_S)$ has the same form as in the SM with the change $\Phi\to\Phi_S$.

In this work we consider the Inert Doublet Model (IDM) which is a 2HDM with $Z_2$-symmetric Lagrangian and a $Z_2$-symmetric vacuum state. Exact $Z_2$ symmetry in the model provides us a Dark Matter candidate. We consider the different variants of the evolution of the Universe after inflation, that lead towards the Inert phase today.

\paragraph{$Z_2$ symmetry} On this potential one may impose two discrete symmetries of $Z_2$ type, called here $D$ and $S$-symmetry \cite{GKKS10}. The corresponding transformations of scalar doublets are given by:
\begin{eqnarray}
 S: \quad	\Phi_{S} \xrightarrow{S} -\Phi_{S},\quad
	\Phi_{D} \xrightarrow{S} \Phi_{D},\qquad
D: \quad \Phi_{S} \xrightarrow{D} \Phi_{S},\quad
	\Phi_{D} \xrightarrow{D} -\Phi_{D},
	\label{dtransf}
\end{eqnarray}
while SM fields are even under both $S$ and $D$ transformation. 

We will request the explicit $Z_2$-symmetry ($D$ or $S$-symmetry) in the potential ($\lambda_6 = \lambda_7 = m_{12}^2 =0$), and without loss of generality one can set $0 > \lambda_5 \in \mathbb{R}$. $Z_2$ symmetric potential has then 7 free parameters $m_{11}^2, m_{22}^2, \lambda_{1-5}$. Those symmetries may be spontaneously violated by the non-vanishing v.e.v of one or two doublets $\langle\Phi_{S}\rangle,\langle\Phi_{D}\rangle$. 
%
%
Note, that the Yukawa term violates $S$-symmetry
even if
$\langle\Phi_{S}\rangle=\langle\Phi_{D}\rangle=0$, while it respects $D$-symmetry in
any order of perturbation theory.

\paragraph{Positivity conditions} The existence of the stable vacuum is guaranteed by the positivity conditions imposed od $V$. They assure that the potential is bounded from below, and thus the extremum with the lowest energy is be the global minimum of the potential (vacuum). The positivity constrains relevant for this analysis are:
\begin{eqnarray}
& \lambda_1>0\,, \quad \lambda_2>0, \quad R + 1 >0 \quad \textrm{ with }  \quad R = \lambda_{345}/\sqrt{\lambda_1 \lambda_2}, \quad \lambda_{345}=\lambda_3+\lambda_4+\lambda_5. \label{Rdef} &
\end{eqnarray}

\paragraph{Vacua in $Z_2$ symmetric 2HDM} The general solution of the extremum conditions of the potential $V$ is:
\begin{eqnarray}
        \langle\Phi_S\rangle =\dfrac{1}{\sqrt{2}}\left(\begin{array}{c} 0\\
        v_S\end{array}\right),\quad \langle\Phi_D\rangle
        =\dfrac{1}{\sqrt{2}}\left(\begin{array}{c} u \\ v_D
        \end{array}\right), \quad (v^2=v_S^2+|v_D^2|+u^2). 
\end{eqnarray}
Depending on the values of the $v_S, v_D, u$ parameters, different extrema can be realized in the model. Below we list the basic properties of possible extrema provided they are realized as the vacua. 

\emph{The Charge breaking vacuum} $Ch$ with $u \not = 0$ and $v_D=0$ leads to the electric charge non-conservation, as $U(1)_{EM}$ symmetry is broken. 

\emph{The electroweak symmetric vacuum} $EW\! s$ corresponds to $u = v_S = v_D = 0$ and $m_{11,22}^2 <0$. Gauge bosons and fermions are massless, while the doublets have non-zero masses $\frac{|m_{11}^2|}{\sqrt{2}}$ and $\frac{|m_{22}^2|}{\sqrt{2}}$, respectively.

\emph{The Inert vacuum} $I_1$ requires $u = v_D = 0$ and $v_S^2 = v^2 = m_{11}^2/\lambda_1>0$. Fermions and gauge bosons are massive, the scalar sector contains SM-like Higgs $h_S$ and dark scalars $D_H, D_A, D^\pm, D_H$. This is the only $D$-symmetric vacuum that can provide the DM candidate (more details in the next section) \cite{Deshpande:1977rw,Barbieri:2006dq}.

\emph{The inertlike vacuum} $I_2$  with $u = v_S = 0$ and $v_D^2 = v^2 = m_{22}^2/\lambda_2>0$ is "mirror-symmetric" to the inert vacuum
$I_1$, with  one Higgs particle $h_D$ and  four scalar particles: $S_H,\,S_A,\,S^\pm$. Here we have no DM candidate, as $S$-symmetry is violated by the Yukawa term ($S$-scalars interact with fermions) and $D$-symmetry is spontaneously violated by the vacuum state. Note that all fermions, by definition interacting only with $\Phi_S$ with vanishing v.e.v.
 $\langle\Phi_S\rangle =0$,  are massless.

\emph{The mixed vacuum} $M$ with $u = 0, \, v_S^2=\frac{m_{11}^2\lambda_2-\lambda_{345}m_{22}^2}{\lambda_1\lambda_2-\lambda_{345}^2}>0,\, v_D^2=\frac{m_{22}^2\lambda_1-\lambda_{345}m_{11}^2}{\lambda_1\lambda_2-\lambda_{345}^2}>0,$ violates the full $Z_2$ symmetry of the potential. There are massive gauge bosons and fermions.  Five massive Higgs bosons exist: two charged $H^\pm$ and three neutral ones, the CP-even $h$ and $H$ and CP-odd  $A$. 

The realization of different types of vacua depends on the value of the parameters of the potential. $M$ and $Ch$ vacua can be realized in the separate regions of $(\lambda_4,\lambda_5)$ parameter space, while $I_1$ (or $I_2$) can overlap $M$ and $Ch$ \cite{Blois}. Note, that in the region where $Ch$ can be realized, the lightest dark scalar is $D^{\pm}$.

%

\section{Inert Doublet Model and dark matter}

If the $I_1$ extremum realizes a vacuum then Universe is descibed by the IDM. It predicts the existence of four dark scalars $D_H,\,D_A,
D^\pm$ and the Higgs particle $h_S$, which is interacting
with the fermions and gauge bosons just as the Higgs boson in the SM.

Inert state  is invariant under the $D$-transformation just as the whole Lagrangian \eqref{lagrbas}. Therefore, the  $D$-parity is conserved and  due to this fact
the lightest $D$-odd particle is stable, being a good  DM candidate.
Masses of the scalar particles are:
\begin{eqnarray}
M_{h_s}^2=\lambda_1v^2 \,,\quad M_{D^\pm}^2=(\lambda_3 v^2-m_{22}^2)/2\,, \quad
M_{D_H,D_A}^2=M_{D^\pm}^2+(\lambda_4 \pm \lambda_5)v^2/2\,. 
\end{eqnarray}
These masses can be used to express the parameters of the potential $V$ after EWSB. Since after EWSB the potential has 6 free parameters, one also needs two self-couplings to describe the model.  
\textit{Triple and quartic} couplings between SM-like Higgs $h_S$ and DM candidate $D_H$, i.e. $D_H D_H h_S$ and $D_H D_H h_S h_S$, are proportional to $\lambda_{345}$. $\lambda_2$ is related only to \emph{quartic} self-couplings, e.g. $D_H D_H D_H D_H$. The remaining self-coupling, $\lambda_3$, governs the $D^\pm$ interactions like $D^+ D^- h_S$ and $D^+ D^- h_S h_S$ vertices.

The value of $\lambda_{345}$ strongly affects the DM interactions relevant for $\Omega_{DM} h^2$, the energy relict density of DM. This is because it governs the main decay channel in wide region of parameter space ($D_H D_H$ annihilation into fermions via $h_S$ exchange).
The value of $\lambda_2$ parameter does not influence $\Omega_{DM} h^2$ explicitly.

\paragraph{Collider constraints on scalars' masses} Strong limitations for the physics beyond SM come from the electroweak precision tests. For IDM both light and heavy Higgs particle is allowed \cite{Barbieri:2006dq}. EWPT constrain the mass splittings between the dark particles $\delta_A = M_{D_A} - M_{D_H}, \, \delta_{\pm} = M_{D^\pm} - M_{D_H}$ \cite{Dolle:2009fn}. If SM Higgs is heavy, then large $\delta_{\pm}$ is needed, while $\delta_{A} $ could be small.  For a light Higgs boson, the allowed region corresponds to  $\delta_{\pm} \sim \delta_{A} $ with mass splittings that could be large.


MSSM constraints from LEP II were interpreted within the IDM in \cite{Lundstrom:2008ai}. This analysis excludes the following region of masses: $M_{D_H} < 80$ GeV, $M_{D_A} < 100$ GeV and $\delta_A > 8$ GeV. For $\delta_A <8$ GeV the LEP I limit $M_{D_H} + M_{D_A} > M_Z$ applies.



\paragraph{DM relict density constraints} 

%
%
%
Various studies \cite{Dolle:2009fn,LopezHonorez:2006gr,Hambye:2007vf,Honorez:2010re} show that for IDM there are three allowed regions of $M_{D_H}$, which may give $\Omega_{DM}h^2$ in agreement with the astrophysical estimations $\Omega_{DM}h^2=0.112 \pm 0.009$ \cite{PDG}. Those regions are: (i) light DM particles with mass close to and below $10 \textrm{ GeV}$, (ii) medium mass regime of $40-80 \textrm{ GeV}$ and (iii) heavy DM of mass larger than $500 \textrm{ GeV}$. 

Astrophysical estimations of $\Omega_{DM} h^2$ may be used to give the limitations for $|\lambda_{345}|$ depending on the chosen value of masses of $D_H$ and other scalars \cite{Dolle:2009fn,LopezHonorez:2006gr}, however not on coupling $\lambda_2$.

\section{Thermal evolution of the Universe}\label{evolution}

To study the earlier history of the Universe after inflation we consider thermal evolution of the Lagrangian in the   first nontrivial approximation
\cite{iv2008,GIK09,GKKS10}. In this approximation the Yukawa couplings and  the quartic coefficients $\lambda's$  remain unchanged, while the mass parameters $m_{ii}^2 \; (i=1,2)$ vary with temperature $T$ as follows:
$m_{ii}^2(T) = m_{ii}^2-c_iT^2, \, c_1 = c_1(\lambda) + c(g,g^{\prime}) + c_1(g_t,g_b), \, c_2 = c_2(\lambda) + c(g,g^{\prime})$. 
%
%
Here $c_i(\lambda) = (3 \lambda_i + 2 \lambda_3 + \lambda_4)/6 $ are the scalar corrections; $c(g,g^\prime)$ is the contribution from the EW gauge couplings, which is the same for both $c_1$ and $c_2$. Only $c_1$ receives $c_1(g_t,g_b)$ correction, that comes from the interaction of $\Phi_S$ with $t$ amd $b$ quarks.

%
%

In virtue of positivity conditions the sum of evolution coefficients is positive: $c_2+c_1>0$. For $R>0$ both $c_i>0$, while for $R<0$ arbitrary signs of $c_{1,2}$ are possible. The restoration of of EW symmetry for high $T$ \cite{Gavela:1998ux} requires positive $c_1,c_2$ \cite{GKKS10}. Here we limit ourselves to the neutral vacua and the case of restoration of EW symmetry.

\begin{figure}
\includegraphics[scale=0.4]{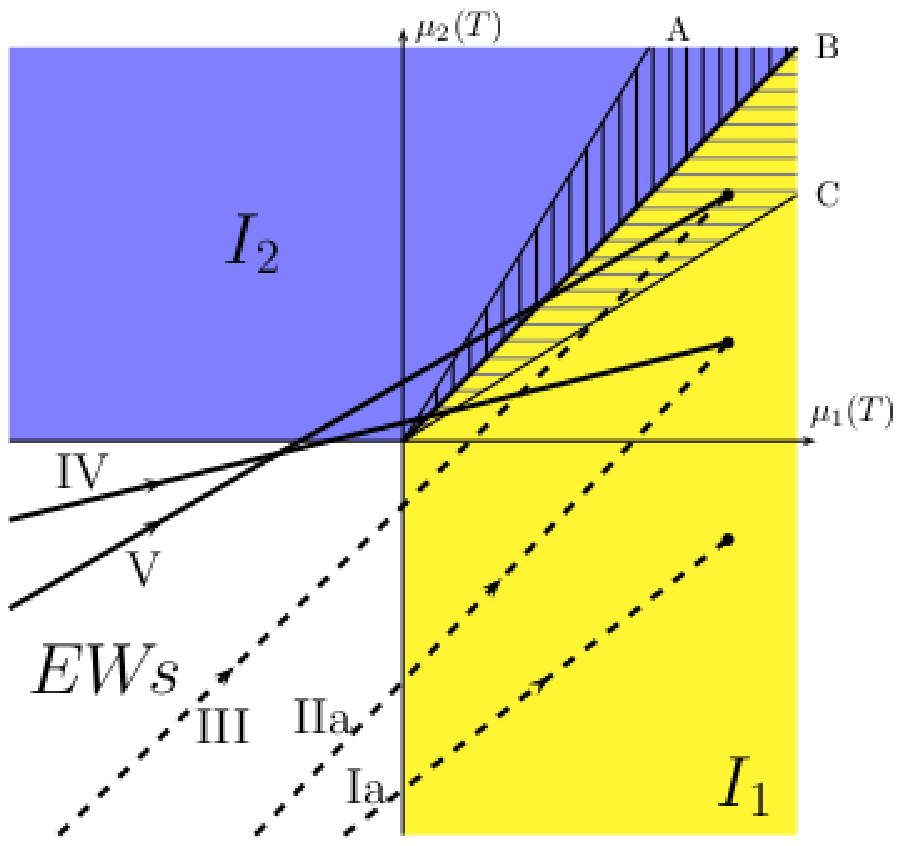}~a)
\hfil
\includegraphics[scale=0.4]{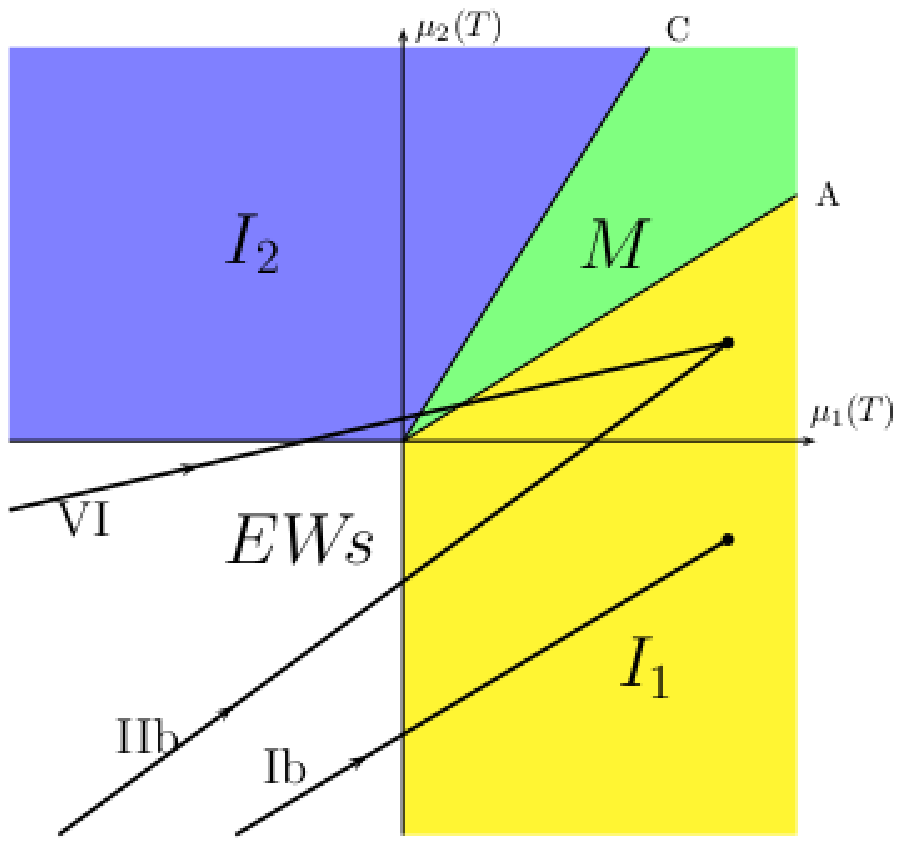}~b)
\hfil
\includegraphics[scale=0.4]{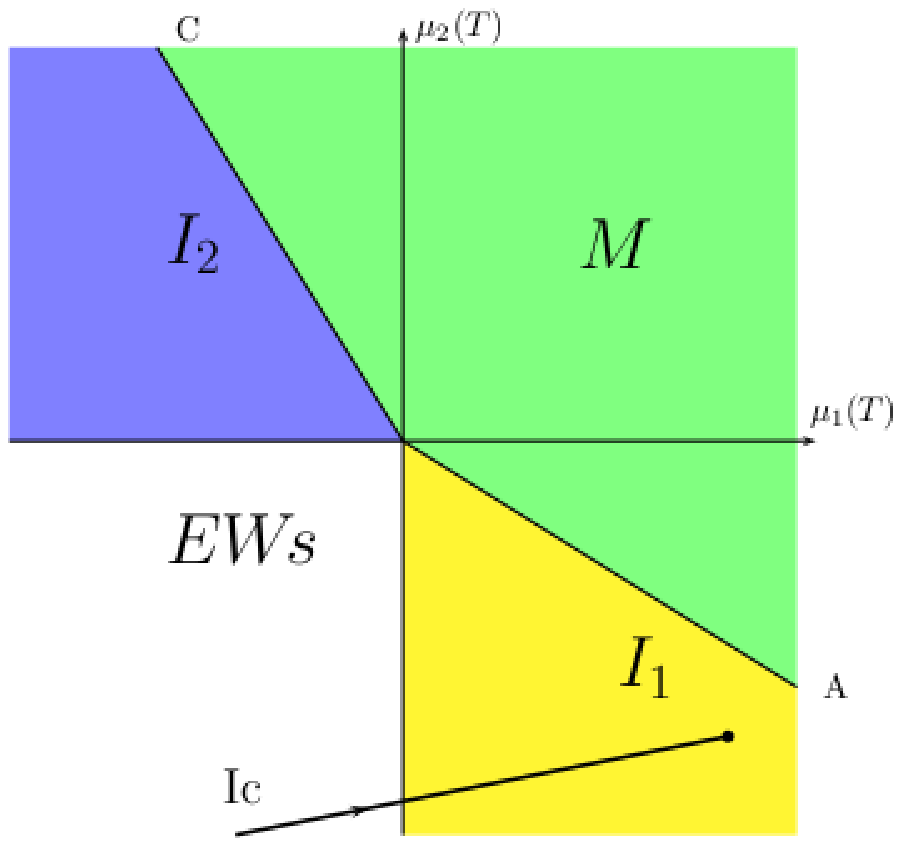}~c)
\caption{Possible vacua and evolution to a current states (dots) represended by rays on $(\mu_1,\mu_2)$ plane for \textbf{a)} $R>1$, \textbf{b)} $1>R>0$, \textbf{c)} $
0>R>-1$. The boundary lines are $A:\mu_2 = \mu_1 R$,\, $B:\mu_2 = \mu_1$,
$\, C:\mu_2 = \mu_1/ R$. Blue (dark shade) region represents $I_2$ vacuum,
yellow (light shade) region -- $I_1$ vacuum and green (medium shade)
region -- $M$ vacuum. In the hatched regions between lines $A,B$ and $B,C$ $I_1$ and $I_2$
minima co-exist.}
\label{fig:evol}
\end{figure}

As the Universe is cooling down the potential $V$, with temperature dependent quadratic coefficients, may have different ground states, discussed in sec.1. Figures \ref{fig:evol}a-c present the types of evolution for different ranges of $R$. In case of rays Ia-c, Ia-b, III there is a single phase transition $EW\! s \to I_1$. For $R>1$ there is a unique possibility of the 1st-order phase transition between $I_2$ and $I_1$ vacua (rays IV and V). Also in this case we have the possibility of co-existence of the vacuum $I_1$ and local minimum $I_2$ for $T=0$ (rays III and V) \cite{Sokolowska:2011yi}. For $0<R<1$ also $M$ can be a vacuum (see \cite{GKKS10}) and for ray VI Universe goes through a sequence of three 2nd-order phase transitions $EW\! s \to I_2 \to M \to I_1$. If $R<0$ the only possible ray that corresponds to the EW symmetry in the past is ray Ic. 

%
%

\begin{figure}
\includegraphics[scale=0.8]{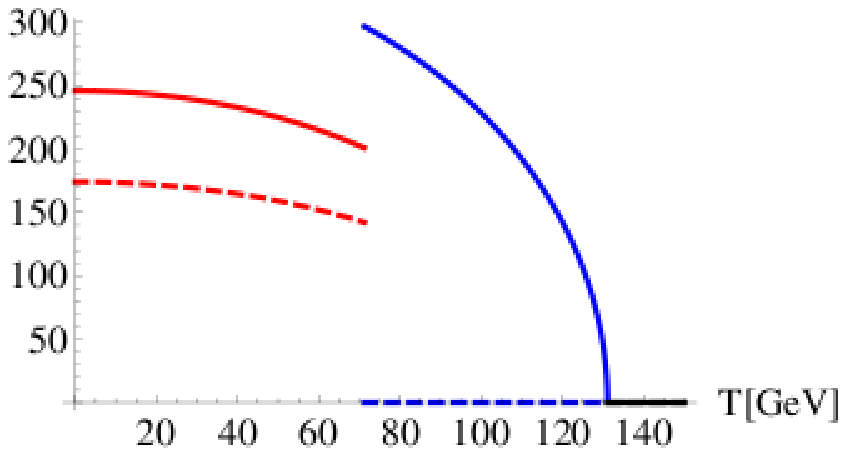}~a)
\hfil
\includegraphics[scale=0.8]{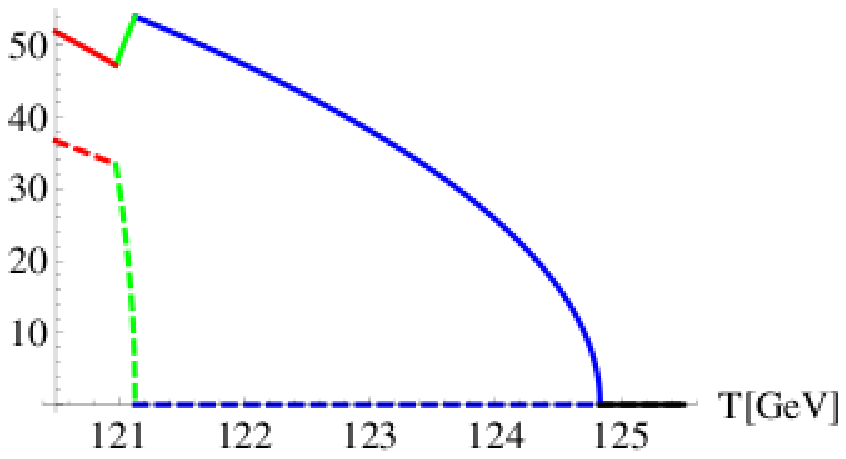}~b)
\caption{Evolution of v.e.v $v(T)$ (solid line) and top mass $m_t(T)$ (dashed line) for example of \textbf{a)} ray V with $\lambda_2 = 0.05$, \textbf{b)} ray VI with $\lambda_2 = 0.125$ for $M_{h_S} = 120 \textrm{ GeV}, M_{D_H} = 45 \textrm{ GeV}, M_{D_A} = 115 \textrm{ GeV}, M_{D^\pm} = 125 \textrm{ GeV}, \lambda_{345} = 0.17$. Parameters were chosen to fulfill the existing collider contraints \cite{Sokolowska:2011yi}.}
\label{fig:examples}
\end{figure}

Figure \ref{fig:examples} presents the temperature evolution of v.e.v $v$, proportional to the $M_W$ mass, and the top quark mass $m_t$ in two types of sequences represented by ray V (fig.\ref{fig:examples}a) and ray VI (fig.\ref{fig:examples}b).  In the first picture the effects of the 1st-order phase transition are visible as a discontinuity during $I_2 \to I_1$. Second sequence consists of the three 2nd-order phase transitions.


\paragraph{Conclusions} IDM is a model which may provide the DM candidate in agreement with WMAP obserwations. We argue, that during thermal evolution the Universe can pass through various intermediate phases, before it reaches the Inert phase. In those intermediate phases there is no DM candidate and gauge bosons and fermions may have different masses than in the Inert phase.

\paragraph{Acknowledgement}
We are thankful to I. F. Ginzburg and K. A. Kanishev for cooperation. We would like to thank the organizers of the School, especially George Zoupanos for very nice scientific atmosphere.
Work was partly supported by Polish Ministry of Science and Higher Education Grant N N202 230337 and 
EU Marie Curie Research Training Network HEPTOOLS, under contract MRTN-CT-2006-035505.



\begin{thebibliography}{10}

\bibitem{Lee:1973iz}
  T.~D.~Lee,
  Phys.\ Rev.\  D {\bf 8} (1973) 1226.

\bibitem{GKKS10}
I. F. Ginzburg, K.A. Kanishev, M. Krawczyk, D. Sokolowska, Phys.\ Rev.\ D {\bf 82} (2010) 123533.

\bibitem{Deshpande:1977rw}
  N.~G.~Deshpande and E.~Ma,
  Phys.\ Rev.\  D {\bf 18} (1978) 2574;

\bibitem{Barbieri:2006dq}
  R.~Barbieri, L.~J.~Hall and V.~S.~Rychkov,
  Phys.\ Rev.\  D {\bf 74} (2006) 015007.


\bibitem{Blois}
  M.~Krawczyk and D.~Soko\l owska, Proceedings of 21st Rencontres de Blois, 
  arXiv:0911.2457 [hep-ph].

\bibitem{Dolle:2009fn}
  E.~M.~Dolle and S.~Su,
  Phys.\ Rev.\  D {\bf 80} (2009) 055012.


\bibitem{Lundstrom:2008ai}
  E.~Lundstrom, M.~Gustafsson and J.~Edsjo,
  Phys.\ Rev.\  D {\bf 79} (2009) 035013.

\bibitem{LopezHonorez:2006gr}
  L.~Lopez Honorez, E.~Nezri, J.~F.~Oliver and M.~H.~G.~Tytgat,
  JCAP {\bf 0702} (2007) 028.
\bibitem{Hambye:2007vf}
  T.~Hambye and M.~H.~G.~Tytgat,
  Phys.\ Lett.\  B {\bf 659} (2008) 651.
\bibitem{Honorez:2010re}
  L.~L.~Honorez and C.~E.~Yaguna,
  JHEP {\bf 09} (2010) 046.


\bibitem{PDG} Particle Data Group. {\it Journ. of Phys.} {\bf G 37} \#7A (2010) 075021

\bibitem{iv2008}
  I.~P.~Ivanov,
  Acta Phys.\ Polon.\  B {\bf 40} (2009) 2789.


\bibitem{GIK09}
  I.~F.~Ginzburg, I.~P.~Ivanov and K.~A.~Kanishev,
  Phys.\ Rev.\  D {\bf 81} (2010) 085031.




\bibitem{Gavela:1998ux}
  M.~B.~Gavela, O.~Pene, N.~Rius and S.~Vargas-Castrillon,
  Phys.\ Rev.\  D {\bf 59} (1999) 025008.
  



\bibitem{Sokolowska:2011yi}
D. Soko\l owska, arXiv:1104.3326 [hep-ph].


\end{thebibliography}
\end{document}